\documentclass[twocolumn]{aa} 
\usepackage{amsmath}

\usepackage{cool} 
\usepackage{bm} 

\usepackage{tabularx}
\usepackage{booktabs}
\usepackage{graphicx}
\usepackage{natbib,twoopt}

\usepackage{hyperref}                
\usepackage{breakurl}

\usepackage{savesym}
\savesymbol{tablenum}
\usepackage{siunitx}
\restoresymbol{SIX}{tablenum}

\DeclareRobustCommand{\fref}[1]{Figure~\ref{#1}}

\DeclareRobustCommand{\tref}[1]{Table~\ref{#1}}

\bibpunct{(}{)}{;}{a}{}{,}    

\DeclareRobustCommand{\numvortObs}{26,988}

\begin{document}  

\title{Vortex Flows in the Solar Atmosphere: Automated Identification and Statistical Analysis}

\author{I. Giagkiozis\inst{1}
\and
V. Fedun\inst{2}
\and 
E. Scullion\inst{3}
\and 
G. Verth\inst{1}}

\institute{School of Mathematics and Statistics, University of Sheffield, Hicks Building, Hounsfield Road, Sheffield, S3 7RH, UK
\and
Automatic Control and Systems Engineering Department, University of Sheffield, Amy Johnson Building, Mappin Road, S1 3JD, UK
\and
Department of Mathematics \& Information Sciences, Northumbria University, Newcastle upon Tyne, NE1 8ST, UK}

\abstract
   {}
   {Due to the fundamental importance of vortices on the photosphere, in this work we aim to fully automate the process of intensity vortex identification to facilitate a more robust statistical analysis of their properties.}
   {Using on-disk observational data of the Fe~{\sc i} continuum, the process of vortex identification is fully automated, for the first time in solar physics, with the help of an established method from hydrodynamics initially employed for the study of eddies in turbulent flows \citep{Graftieaux:2001aa}.}
   {We find that the expected lifetime of intensity vortices is much shorter ($\approx 17 \si{\second}$) compared with previously observed magnetic bright point swirls. Our findings suggest that at any time there are $1.4 \times 10^{6}$ such small-scale intensity vortices covering about $2.8\%$ of the total surface of the solar photosphere.}
   {}


\maketitle

\section{Introduction}\label{sec:introduction}

Traditionally photospheric intensity flow fields have been traced using local correlation tracking of (magnetic) bright points and the revealed vortex flows have been identified by eye. This manual approach has two major shortcomings, i) it introduces observational bias into the statistical analysis and ii) a large number of vortex flow fields are most likely missed simply due to the sheer scale of the task, which also has adverse effects on the variance of the statistical analysis. Small-scale vortices in the quiet Sun regions are widely accepted to form due to turbulent convection and the \textit{bath-tub} effect \citep[e.g.][]{Shelyag:2011aa,Kitiashvili:2012aa,Shelyag:2012aa}. Solar photospheric vortex flows have drawn the attention of researchers as they have the potential to excite a wide range of MHD waves, e.g. slow and fast magneto-acoustic as well as Alfv\'{e}n \citep{Fedun:2011ab,Mumford:2015aa,Mumford:2015ab}. Vortex flows also appear to have a prominent role in both direct and alternating current models of solar atmospheric heating. In direct current models, neighboring magnetic flux tubes (or strands) can become magnetically twisted under the influence of photospheric vortices. This, in turn, implies that current sheets may develop at the interface between such strands allowing the possibility of magnetic reconnection \citep{Parker:1972aa,Parker:1983aa,Parker:1983ab,Klimchuk:2015aa}. In alternating current models, photospheric vortices can be seen as MHD wave drivers \citep{Fedun:2011ab,Mumford:2015aa,Mumford:2015ab} and as precursors to large scale solar tornadoes \citep{wedemeyer2012magnetic,Wedemeyer:2013aa,Amari:2015aa}. These tornadoes have an estimated net positive Poynting flux of $440\si{\watt \per \square \metre}$ that is more than adequate to heat the quiet solar atmosphere whose energy flux requirement is estimated to be approximately $300\si{\watt \per \square \metre}$ \citep{Withbroe:1977aa}. Unfortunately, despite the increasing interest in these coherent flows in the solar photosphere, the number of observations reported in the literature is still based on small sample sizes, with reports often associated with only a few detected events \citep[see for example][]{Steiner:2010aa,Palacios:2012aa,Park:2016aa}, or tens of observations \citep[e.g.][]{Bonet:2008:convvortex,Bonet:2010:sunrise,Vargas-Dominguez:2011aa}. 

In this paper we present a fully automated method to identify vortex flows, namely the center of circulation and their flow boundary that is based on local correlation tracking \citep{fisher2008flct} applied to photospheric intensity observations, combined with an established method for identifying vortices used in the study of turbulence \citep{Graftieaux:2001aa}. Subsequently, we estimate characteristic vortex parameters, such as lifetime diameter, mean perpendicular velocity and area. The main results of this paper are the following. There is an abundance of small-scale intensity vortices in the quiet Sun and their typical lifetimes are approximately $17$ seconds. We estimate that at any given time, the expected number of vortices in the photosphere is $1.4 \times 10^{6}$ and that they occupy $2.8\%$ of the photosphere. Although the area of these vortices may appear small in the photosphere, even if only a tenth of these vortex flows reaches the lower corona they may occupy more than $17\%$ of its total area.

\section{Observations and Vortex Identification Process}
\subsection{Observations}\label{sec:observations}

\begin{figure*}
\includegraphics[width=\hsize]{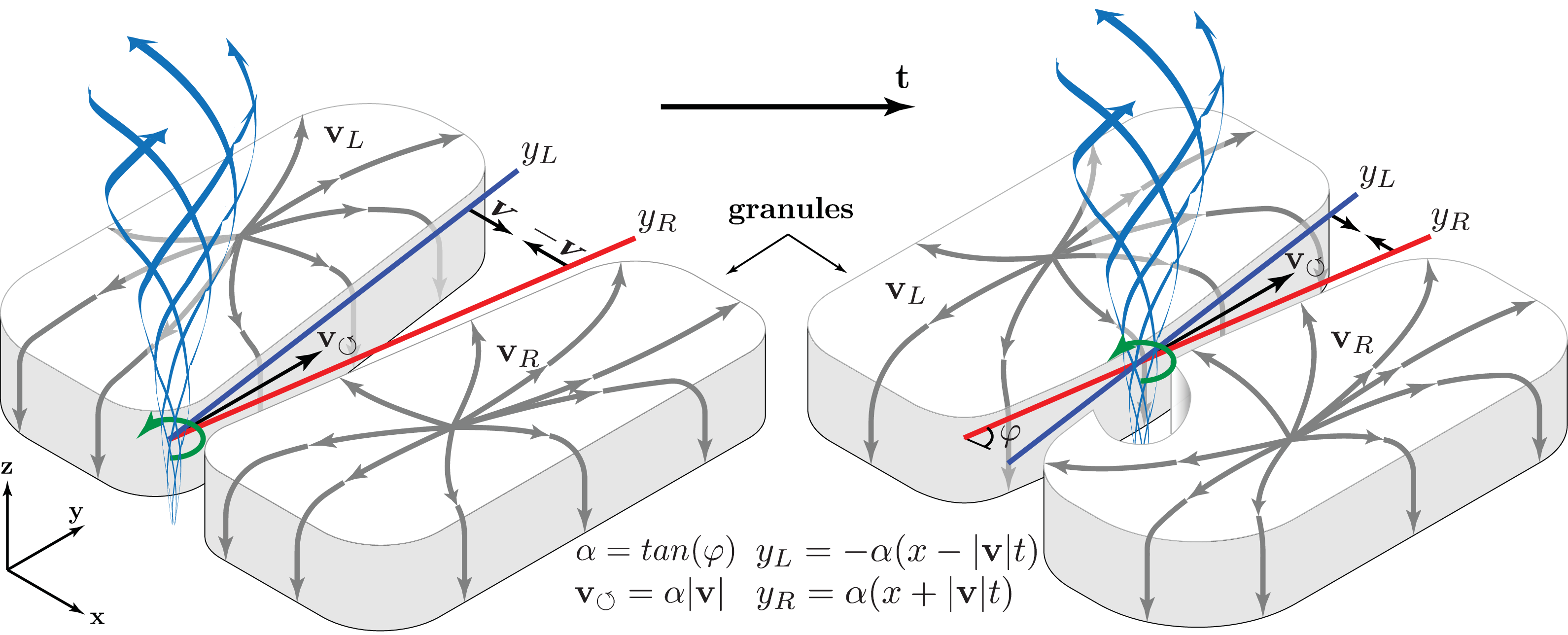}
\caption{Cartoon of the proposed physical mechanism modelling the high velocity of vortex centers. The line segments $y_L$ and $y_R$, shown in blue and red color, respectively, represent the edges of two neighboring granules. In this instance, the two edges are moving towards each other with speed $|\mathbf{v}|$. The streamlines in the plane outline the velocity field near the edges of the granules, with $\mathbf{v}_L$ and $\mathbf{v}_R$ denoting the velocity field in the left and right granule, respectively. The velocity of the vortex center is labeled $\mathbf{v}_{\circlearrowleft}$. The blue streamlines in the z-direction depict magnetic field lines above the vortex center. The black arrow shows the evolution. \label{fig:vfcartoon3d}}
\end{figure*}

\begin{figure*}
\includegraphics[width=\hsize]{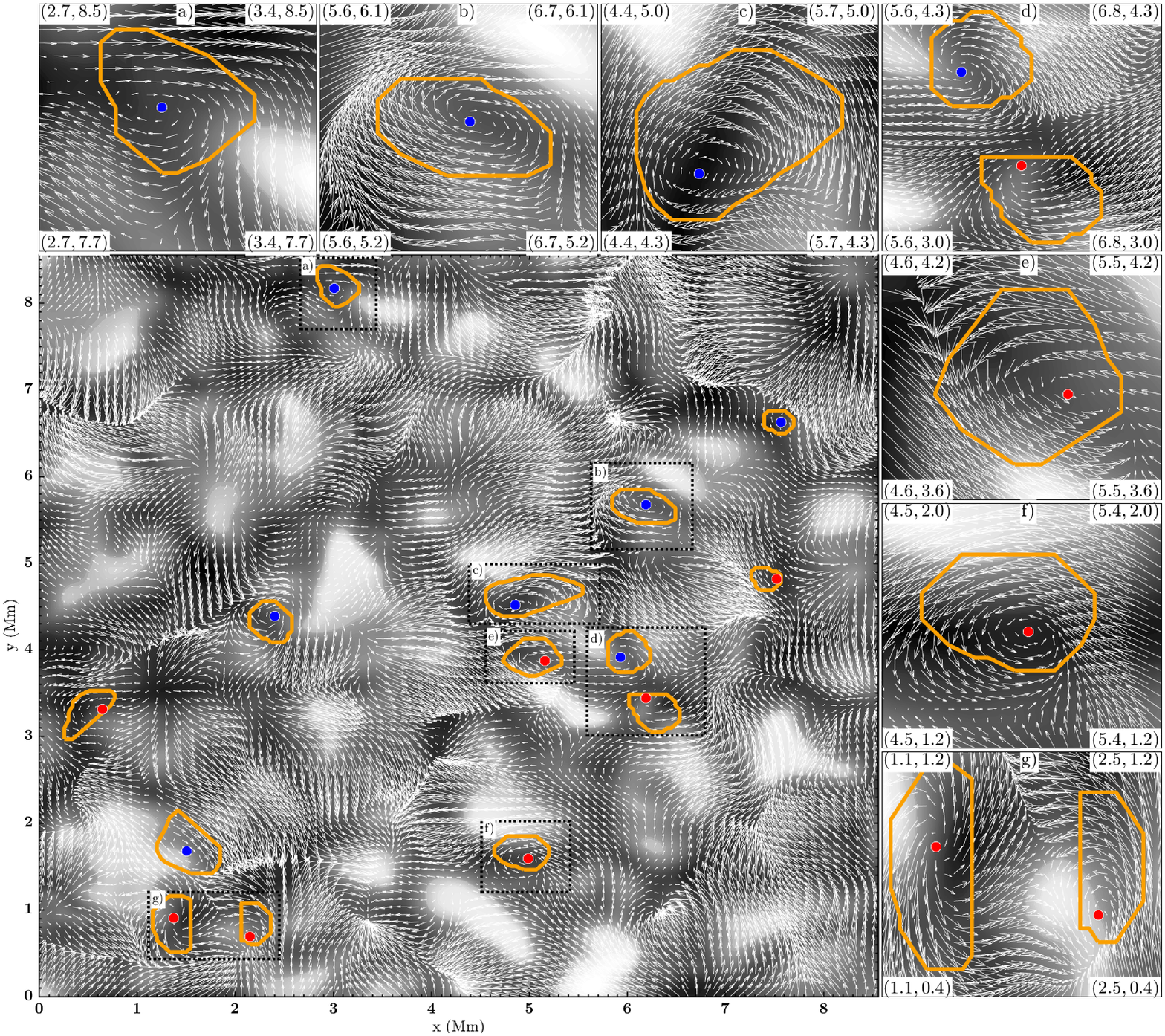}
\caption{A snapshot of the estimated velocity field based on the Fe~{\sc i} continuum (intensity shown in grayscale) using local correlation tracking (LCT), illustrating the identified vortices and their boundaries. The circles denote the vortex center, with red referring to counter clockwise vortices (positive) and blue clockwise vortices (negative). The orange border line denotes the vortex boundary. (See on-line supplementary materials for a movie).\label{fig:vortices}}
\end{figure*}

The observations investigated here were carried out between 08:07:24--09:05:46~UT on the 21st June 2012, with the CRisp Imaging SpectroPolarimeter (CRISP) at the Swedish 1-m Solar Telescope \citep[SST:][]{Scharmer:2003aa,Scharmer:2008aa} on La Palma. The image resolution of the CRISP observations is $0.059^{\prime\prime}$per pixel. A quiet Sun region very close to disk center was observed with an effective Field-Of-View (FOV) of 55$\times$55~arcsec, centered on solar-\textit{x}=$-3.1^{\prime\prime}$ and solar-\textit{y}=$69.9^{\prime\prime}$. The required accuracy of the pointing of the CRISP FOV, in heliocentric coordinates, was achieved through co-alignment of bright-points observed with the CRISP wideband images, together with co-temporal continuum images in $170.0\si{\nano\metre}$ from the Solar Dynamics Observatory / Atmospheric Imaging Assembly \citep[SDO/AIA:][]{Lemen:2012aa}. The spectro-polarimetric sequences have a post-reduction mean cadence of $8.25\si{\second}$. After acquisition, the data was processed with the Multi-Object Multi-Frame Blind Deconvolution (MOMFBD) algorithm \citep{van-Noort:2005aa,van-Noort:2008aa,de-la-Cruz-Rodriguez:2015aa}.

\subsection{Vortex Identification Process}\label{sec:methods}
The automated vortex identification methodology we present splits into four stages: i) pre-processing, ii) velocity field estimation, iii) vortex identification and, iv) vortex lifetime estimation.
The intensity maps obtained from observations have varying intensity at different times that appears to be due to atmospheric effects. These intensity variations are a few standard deviations from the mean and the effect is global. To counter these effects image histogram equalization \citep[e.g.][]{pizer1987adaptive} was used in the following way: 
\begin{itemize}
\item First, the expected distribution of intensities is estimated by means of averaging the histogram distributions across all frames. The rationale for this is that the Sun is not expected to change its general power emission spectrum during the time of the observation. 
\item Once the expected intensity distribution has been obtained, histogram equalization is applied to all frames using that distribution as a reference. 
\end{itemize}
This procedure is fast and efficiently removes inter-frame \textit{flickering}, and, improves the numerical stability of the LCT method. 

The pre-processing stage removes rapid intensity fluctuations, while preserving the relative counts as much as possible. Subsequently, we remove the mild \textit{seeing} effects in our observations using a Gaussian filter. Although it is well known that atmospheric seeing is a nonlinear effect \citep[see e.g.][]{November:1988aa}, given that the seeing conditions were good, this simple averaging method produced similar results to destreching algorithms and is computationally more efficient. To avoid the reduction of the temporal resolution a moving-average Gaussian filter is employed with a $3\text{dB}$ attenuation at quarter the Nyquist frequency $1/(8T)$ where $T$ is the cadence. The velocity field estimation is performed using fast local correlation tracking (FLCT) \citep{fisher2008flct} with a Gaussian apodizing window with a width of $\sigma = 10$ pixels \citep{Louis:2015aa}. 

Subsequently, for the vortex identification, we implement a proven and established method from the study of turbulence in fluid dynamics. Once the velocity field estimates are found we implement the same approach as \cite{Graftieaux:2001aa} to identify the vortex centers and boundaries. \cite{Graftieaux:2001aa} defined two functions $\Gamma_1$ and $\Gamma_2$, for the identification of the vortex centers and boundaries, respectively. The function $\Gamma_1$ used in this work is, 
\begin{equation}\label{eqn:gamma1:disc}
\Gamma_1(\mathbf{x}_p) = \frac{1}{|S|} \sum_{S} \frac{\left((\mathbf{x}_m - \mathbf{x}_p) \times \mathbf{v}_m\right) \cdot \mathbf{1}_z}{||\mathbf{x}_m - \mathbf{x}_p||_2 \cdot ||\mathbf{v}_m||_2}.
\end{equation}
Here, $S = \{ \mathbf{x}_m \, : \, ||\mathbf{x}_m - \mathbf{x}_p||_2 \leq R \}$ is a disk of radius $R$ about the point $\mathbf{x}_p$, $||\cdot||_2$ is the Euclidean norm, $\mathbf{1}_z$ is a unit vector normal to the plane and $|S|$ is the cardinality of $S$. $\Gamma_1$ defines a scalar field and its magnitude achieves a maximum at unity. \cite{Graftieaux:2001aa} shows that this function achieves this maximum when $\mathbf{x}_p$ is at the center of an axisymmetric vortex. However, given that ideal axisymmetric vortices are quite uncommon, the threshold for classifying a point in $S$ as a potential vortex center is reduced to $0.9$, and, the local maximum of these points is classified as the vortex center. For the identification of the vortex boundary, we use the discrete version of $\Gamma_2$, defined as follows,
\begin{equation}\label{eqn:gamma2:disc}
\Gamma_2(\mathbf{x}_p) = \frac{1}{N} \sum_{S} \frac{\left((\mathbf{x}_m - \mathbf{x}_p) \times (\mathbf{v}_m - \bar{\mathbf{v}}_p)\right) \cdot \mathbf{1}_z}{||\mathbf{x}_m - \mathbf{x}_p||_2 \cdot ||\mathbf{v}_m - \bar{\mathbf{v}}_p||_2},
\end{equation}
where $\bar{\mathbf{v}}_p$ is the mean velocity in the neighborhood of the point $\mathbf{x}_p$. It is shown in \cite{Graftieaux:2001aa} that in the inner core of a vortex the magnitude of $\Gamma_2$ is larger than $2/\pi$. Flows with values of $\Gamma_2 < 2/\pi$ are dominated by strain and when $\Gamma_2 = 2/\pi$ we have a pure shear.

Let us now calculate the vortex centers and their boundaries at every time instance, however, we still need to estimate the lifespan of these vortices. For this purpose we assume that the vortex center can move at approximately the sound speed of the photosphere, about $10\si{\kilo \metre \per \second}$ \citep{Nordlund:2009aa}. If the speed of the vortex centers is comparable to the sound speed, this would suggest that the maximum distance a center could traverse from one frame to the next would be $82.5\si{\kilo \metre}$, which is almost $2$ pixels at the spatial resolution of our data. However, at present the vortex formation mechanism has not been clearly established and if such flows in the photosphere are formed as shown in \fref{fig:vfcartoon3d}, the speed of their center may be much larger than the sound speed. What we suggest in \fref{fig:vfcartoon3d} is the following, the edges of the granules are represented as line segments (red and blue line segments in \fref{fig:vfcartoon3d}). We define the points where the vertical component of the velocity transits from being mostly positive, as is on granules, to being negative, as is the case in the inter-granular lanes. Due to the dynamic nature of the granulation pattern on the photosphere, their edges are in constant relative motion with respect to the edges of neighboring granules. This relative motion, when combined with counter streaming flows of two neighboring granules, can drive vortex flows whose centers can move ($\mathbf{v}_{\circlearrowleft}$) at much larger speed compared with the relative speed that generated them (see \fref{fig:vfcartoon3d}). Therefore, using a conservative estimate we assume that vortices that are within a $4$ pixel radius in two consecutive times, are in fact the same vortex.

\section{Results and Statistical Analysis}\label{sec:statistics}

\begin{figure*}[ht]
\centering
\includegraphics[width=\hsize]{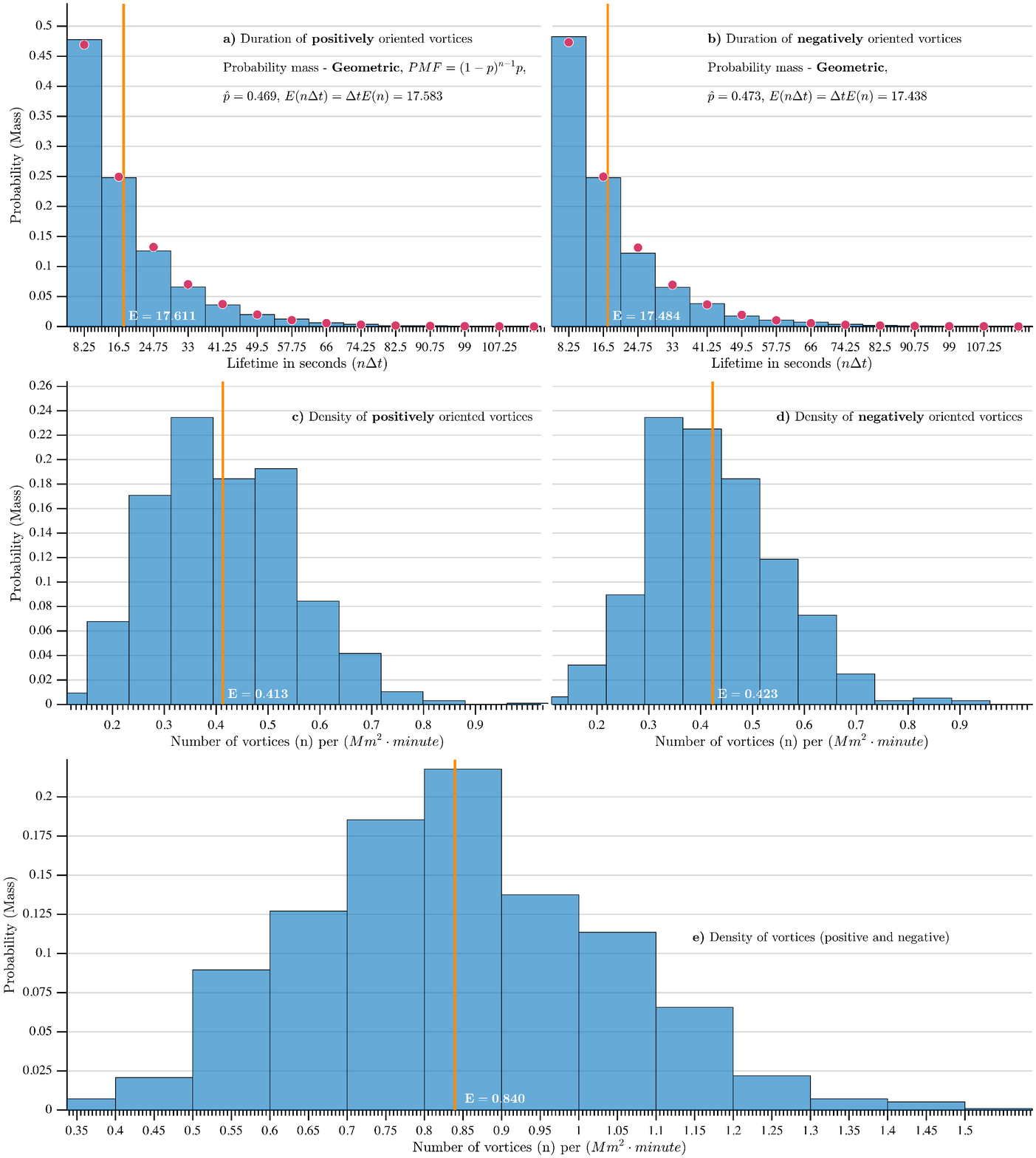}
\caption{Estimates of (\textbf{a} and \textbf{b}) vortex lifetime mass function, and (\textbf{c},\textbf{d} and \textbf{e}) the number of vortices per $Mm^2 \cdot minute$. The red circles denote the best fit of a parametric mass density function (PMF). In this case, the Geometric distribution was a best fit for the lifetimes of the vortices. The orange line, and the white font $E$ on its right, is the expected value calculated from the empirical distribution of the data. Values with a \textit{hat} indicate best fit parameter estimates for the particular distribution, and, $E(\cdot)$ is the expected value. \label{fig:lfncnt:obs}}
\end{figure*}

\begin{figure*}[ht]
\centering
\includegraphics[width=\hsize]{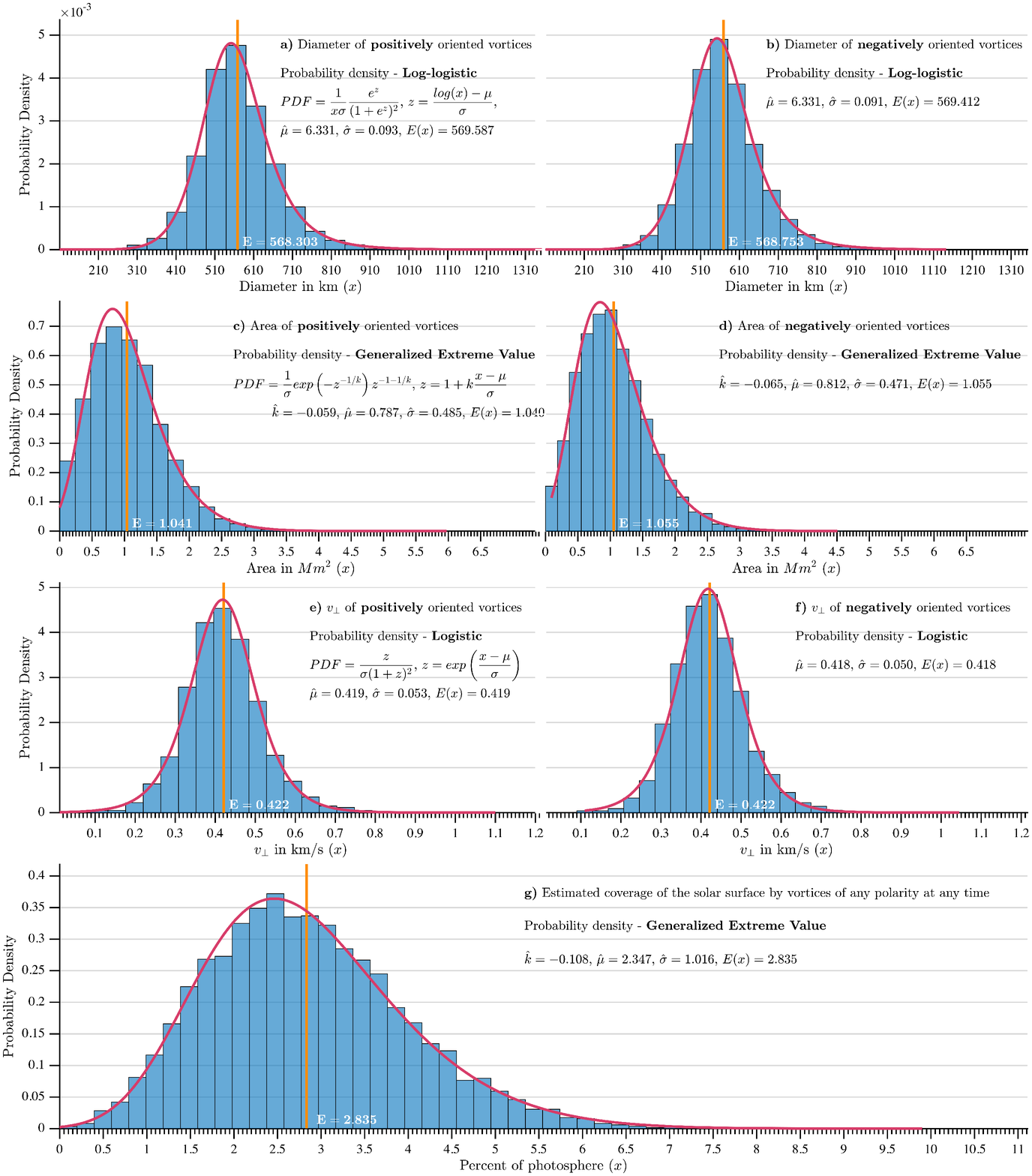}
\caption{Empirical and parametric estimates of the probability density function (PDF) for (\textbf{a} and \textbf{b}) the vortex diameter, which is calculated using the average of the minor and major axis of a best fit ellipse for every vortex, (\textbf{c} and \textbf{d}) the area of vortices (in $Mm^2$), (\textbf{e} and \textbf{f}) the magnitude of the perpendicular velocity ($|v_{\perp}|$) and lastly (\textbf{g}) an estimate of the percentage of the total photosphere covered with intensity vortices. The notation in this figure follows \fref{fig:lfncnt:obs}. \label{fig:ravtota:obs}}
\end{figure*}

A representative example of the results obtained from the vortex identification process is shown in \fref{fig:vortices}. The grayscale denotes intensity, normalized in the range $0$ to $1$ corresponding to black and white, respectively. Over-plotted is the LCT estimate of the \textit{surface} velocity field. The red-filled circles mark counter-clockwise vortex flows (positive), blue circles correspond to clockwise flows (negative) and the orange line delimits the vortex flow boundary. 

\fref{fig:lfncnt:obs} and \fref{fig:ravtota:obs} show statistical results based on a sample size of $N = \numvortObs$ vortices. As the only source of information that is used here is based on LCT applied to intensity observations, we refer to the identified vortex flows as \textit{intensity vortices}. This is to acknowledge line-of-sight integration effects and temperature variations that, from a practical standpoint, lead to the estimated velocity field being a weighted average of the plasma motions at different heights within the spectral line formation height \citep{Nordlund:2009aa}.

We find that the expected lifetime of such vortices is independent of orientation (see a) and b) in \fref{fig:lfncnt:obs}). In fact, we have found no statistically significant deviations in the distributions of positively or negatively oriented vortices for any of the measured parameters, i.e. lifetime, space and time density, diameter, area or perpendicular speed (see \fref{fig:lfncnt:obs} and \ref{fig:ravtota:obs}). What is intriguing however, is that for the majority of vortices (approximately $85\%$) their expected lifetimes are less or equal to three times the cadence ($24.75\si{\second}$). This is much shorter when compared with similar features identified by tracking magnetic bright points (BPs) \citep[e.g.][]{Bonet:2008:convvortex}. The apparent discrepancy could be attributed to errors in LCT, where very short lived structures are the result of errors in the identified velocity field. Notwithstanding this limitation, LCT velocity maps have been shown to be a good first order approximation to the velocity field \citep{Verma:2013aa,Louis:2015aa}. 

Assuming an expected lifetime, for both positive and negative vortices, of $\tau = 0.29 \: \si{\minute}$ (see \textbf{a)-b)} in \fref{fig:lfncnt:obs}) and the space and time density of vortices $d=0.84 \: \si{\per \square \mega \metre \per \minute}$ (see \textbf{e)} in \fref{fig:lfncnt:obs}), we estimate that there will be $\tau \cdot d = 0.244 \: \si{\per \square \mega \metre}$ vortices at any time. In turn, this implies that there are continuously $1.48 \times 10^{6}$ vortices over the whole photosphere. Under the assumption that the expected lifetime, as well as, space and time density in other regions of the photosphere are similar to our observations. If intensity vortices are indeed closely correlated with the actual velocity field, then based on the simulation results reported by \cite{Kitiashvili:2012aa,Kitiashvili:2012ac} we anticipate that their expected size will decrease with the advent of higher spatial resolution observations. A prime example of near future expected capability is the Daniel K. Inouye Solar Telescope (DKIST) whose visible broadband imager is planned to have spatial resolution of $16\si{\kilo \meter}$ to $25\si{\kilo\meter}$ per pixel, at $430.4\si{\nano\meter}$ and $656.3\si{\nano\meter}$, respectively, and cadence of $3.2\si{\second}$ \citep{Berger:2013aa}. 

\fref{fig:ravtota:obs}, panels \textbf{e)} and \textbf{f)} show the distribution of the average perpendicular speed within the vortex boundary. This is calculated by projecting the velocity vector at every point within the vortex to a vector perpendicular to the ray emanating from the vortex center. Lastly, panel \textbf{g)} in \fref{fig:ravtota:obs} provides an estimate of the percent of the area of the photosphere covered by intensity vortices at any time. 

\section{Discussion and Conclusions}\label{sec:conclusions}
\begin{table*}
\begin{center}
\resizebox{2\columnwidth}{!}{%
\begin{tabular}{lrrrrr}
\toprule
	& $d \; (\si{\per \square \mega \metre \per \minute})$ & $\tau \; (\si{\minute})$ & $\tau \cdot d \; (\si{\per \square \mega \metre})$ & $E(\#)$ on Photosphere & Sample size\\ 
\midrule
	Present work & $0.84$ & $0.29$  & $0.244$  & $1.48 \times 10^{6}$ & $\numvortObs$  \\
	\citep{Bonet:2008:convvortex} & $1.8 \times 10^{-3}$ & $5.1$ & $0.92 \times 10^{-2}$ & $0.55 \times 10^{5}$ & $138$ \\ 
	\citep{Bonet:2010:sunrise} & $3.1\times10^{-3}$ & $7.1$ & $2.2 \times 10^{-2}$ & $1.3 \times 10^{5}$ & $42$\\
	\citep{Vargas-Dominguez:2011aa} & $1.6 \times 10^{-3}$ & $15$ & $2.4 \times 10^{-2}$ & $1.46 \times 10^{5}$ & $144$ \\
	\citep{wedemeyer2012magnetic} & ($1.4 \times 10^{-4}$) & ($12.7$) & ($1.8 \times 10^{-3}$) & ($1.1 \times 10^{4}$) & ($14$) \\
\bottomrule
\end{tabular}%
}
\caption{Summary statistics of space and time density ($d$), expected lifetime ($\tau$), number of vortices per $\si{\square \mega \metre}$ ($\tau \cdot d$) and expected number of vortices on the solar photosphere ($E(\#)$). In the last column we also provide the number of vortices on which the statistical analysis is based on. The values in the last row in the table are in parentheses since they do not correspond to observations in the photosphere, however, these are included for reference.\label{table:summary}}
\end{center}
\end{table*}

Vortex flows in the solar atmosphere may contribute significantly to the energy flux requirements for heating the quiet Sun atmosphere. However, for that connection to be established strong evidence is required: i) vortex flows motions are ubiquitous in the solar atmosphere, ii) that these motions appear at different heights, e.g. photosphere, chromosphere and corona. We have shown, that the automated identification approach described in this work results in a significantly larger number of identified vortices compared with previous observational studies. This is evidence consolidating the fact that small-scale vortices are prevalent in the solar photosphere. Most interestingly, an overwhelming majority of these vortices have lifetimes that are often much shorter than previously believed, which suggests that these flows are highly dynamic in nature. 

Due to the episodic nature of the formation of these small-scale vortices, any magnetic field through them will be supplied with a broadband impulse comprised of both torsional and radial components which will generate propagating MHD waves. The presence of a magnetic field in vortices is consistent if we recall that their location is in the inter-granular lanes where the magnetic field concentrations are highest. Both observational and numerical simulations \citep[e.g.][]{Fedun:2011ab,Mumford:2015aa,Mumford:2015ab} support the idea that MHD waves with a broad frequency range can be generated by vortex flows. However this is to be expected on more fundamental grounds due to a particular duality in frequency space. Namely, localization in time, leads to spread (broadening) in the frequency domain and vice versa. In physical terms this implies that the rapidity of vortex formation alongside with deviation from axi-symmetry offer a wave driver that results in waves of different frequencies, albeit with different amplitudes. Regarding energy transport to the upper layers of the atmosphere, numerical simulations suggest that vortex driven MHD waves \citep{Amari:2015aa} are a feasible mechanism.

The most compelling differences compared with previous reports \citep[e.g.][]{Bonet:2008:convvortex,Bonet:2010:sunrise,Vargas-Dominguez:2011aa} are in the expected lifetime and space and time density. \tref{table:summary} shows summary statistics comparing the main results in this work with previous studies that are based on more than 3-4 observed vortices. In our view, there are at least two explanations for this mismatch. First, vortex flows and any type of feature tracking in observations, is time consuming and error prone when performed manually. This increases the likelihood of bias and increased variance. Also, our estimate of lifetimes relies on the accuracy of LCT for the surface velocity field identification, which although has been shown to have reasonable correlation with the true velocity field \citep{Louis:2015aa}, is only a first approximation to small-scale motions in inter-granular lanes. Notwithstanding this uncertainty, given that our chosen automated technique is straightforward to implement, the results can be cross-validated by other studies, which, in our view, is extremely important.

The similarities in scale and location, of vortex flows identified in this work with small-scale whirlpools (or swirls) reported by \cite{Bonet:2008:convvortex,Bonet:2010:sunrise} and \cite{Vargas-Dominguez:2011aa}, lead us to conjecture that these are, in fact, the same flow features in the quiet Sun. If that is indeed the case, then remarkably the number of vortices in the photosphere would be an order of magnitude larger than previous estimates.

The results in this work could suggest that previously identified magnetic tornadoes by 
\cite{wedemeyer2012magnetic}, may also be more numerous by an order of magnitude. If intensity vortices prove to be the root cause of solar tornadoes \citep{Amari:2015aa}, then this would suggest that $10\%$ of the photospheric vortices reach the lower corona forming magnetic tornadoes. This has the extraordinary implication that at least $17\%$ of the area of the lower corona is constantly supplied with a positive Poynting flux of $440 \si{\watt \per \square \metre}$, as opposed to $1.2\%$ implied from \cite{wedemeyer2012magnetic}. The assumptions in this estimate are that the photospheric vortices that do extend up to the lower corona have a mean radius of $1.5\si{\mega \metre}$, their \textit{corrected} expected number is $1.48 \times 10^5$ vortices at every time, i.e. $10\%$ of our intensity vortices on the photosphere, instead of $1.1 \times 10^4$ (see \tref{table:summary}) and that the average net positive Poynting flux in each magnetic tornado is $440 \si{\watt \per \square \metre}$. These implications may be exciting, however, at present we are far from establishing a link between intensity vortices and magnetic tornadoes. Nevertheless, this would be an interesting direction of future enquiry as this could be key in resolving the quiet Sun heating problem.

\begin{acknowledgements}
I.G. would like to thank the Faculty of Science in the University of Sheffield for the SHINE studentship. IG, VF and GV thank the STFC and the Royal Society for the support received. VF and GV, acknowledge the 2016 Sheffield International Mobility Scheme. All the authors would like to thank Matthias Rempel for the engaging discussions on this work. The Swedish 1-m Solar Telescope is operated on the island of La Palma by the Institute for Solar Physics of
Stockholm University in the Spanish Observatorio del Roque de los Muchachos of the Instituto de Astrof\'{i}sica de Canarias. The authors wish to acknowledge the DJEI/ DES/ SFI/ HEA Irish Centre for High-End Computing (ICHEC) for the provision of computing facilities and support.
\end{acknowledgements}



\begin{thebibliography}{}
\expandafter\ifx\csname natexlab\endcsname\relax\def\natexlab#1{#1}\fi

\bibitem[{{Amari} {et~al.}(2015){Amari}, {Luciani}, \& {Aly}}]{Amari:2015aa}
{Amari}, T., {Luciani}, J.-F., \& {Aly}, J.-J. 2015, \nat, 522, 188

\bibitem[{{Berger} \& {ATST Science Team}(2013)}]{Berger:2013aa}
{Berger}, T., \& {ATST Science Team}. 2013, in AAS/Solar Physics Division
  Meeting, Vol.~44, AAS/Solar Physics Division Meeting, 400.02

\bibitem[{{Bonet} {et~al.}(2008){Bonet}, {M{\'a}rquez}, {S{\'a}nchez Almeida},
  {Cabello}, \& {Domingo}}]{Bonet:2008:convvortex}
{Bonet}, J.~A., {M{\'a}rquez}, I., {S{\'a}nchez Almeida}, J., {Cabello}, I., \&
  {Domingo}, V. 2008, \apjl, 687, L131

\bibitem[{{Bonet} {et~al.}(2010){Bonet}, {M{\'a}rquez}, {S{\'a}nchez Almeida},
  {Palacios}, {Mart{\'{\i}}nez Pillet}, {Solanki}, {del Toro Iniesta},
  {Domingo}, {Berkefeld}, {Schmidt}, {Gandorfer}, {Barthol}, \&
  {Kn{\"o}lker}}]{Bonet:2010:sunrise}
{Bonet}, J.~A., {M{\'a}rquez}, I., {S{\'a}nchez Almeida}, J., {et~al.} 2010,
  \apjl, 723, L139

\bibitem[{{de la Cruz Rodr{\'{\i}}guez} {et~al.}(2015){de la Cruz
  Rodr{\'{\i}}guez}, {L{\"o}fdahl}, {S{\"u}tterlin}, {Hillberg}, \& {Rouppe van
  der Voort}}]{de-la-Cruz-Rodriguez:2015aa}
{de la Cruz Rodr{\'{\i}}guez}, J., {L{\"o}fdahl}, M.~G., {S{\"u}tterlin}, P.,
  {Hillberg}, T., \& {Rouppe van der Voort}, L. 2015, \aap, 573, A40

\bibitem[{{Fedun} {et~al.}(2011){Fedun}, {Shelyag}, {Verth}, {Mathioudakis}, \&
  {Erd{\'e}lyi}}]{Fedun:2011ab}
{Fedun}, V., {Shelyag}, S., {Verth}, G., {Mathioudakis}, M., \& {Erd{\'e}lyi},
  R. 2011, Annales Geophysicae, 29, 1029

\bibitem[{Fisher \& Welsch(2008)}]{fisher2008flct}
Fisher, G.~H., \& Welsch, B.~T. 2008, in Subsurface and Atmospheric Influences
  on Solar Activity, Vol. 383, 373

\bibitem[{{Graftieaux} {et~al.}(2001){Graftieaux}, {Michard}, \&
  {Grosjean}}]{Graftieaux:2001aa}
{Graftieaux}, L., {Michard}, M., \& {Grosjean}, N. 2001, Measurement Science
  and Technology, 12, 1422

\bibitem[{{Kitiashvili} {et~al.}(2012{\natexlab{a}}){Kitiashvili},
  {Kosovichev}, {Mansour}, {Lele}, \& {Wray}}]{Kitiashvili:2012aa}
{Kitiashvili}, I.~N., {Kosovichev}, A.~G., {Mansour}, N.~N., {Lele}, S.~K., \&
  {Wray}, A.~A. 2012{\natexlab{a}}, \physscr, 86, 018403

\bibitem[{{Kitiashvili} {et~al.}(2012{\natexlab{b}}){Kitiashvili},
  {Kosovichev}, {Mansour}, \& {Wray}}]{Kitiashvili:2012ac}
{Kitiashvili}, I.~N., {Kosovichev}, A.~G., {Mansour}, N.~N., \& {Wray}, A.~A.
  2012{\natexlab{b}}, \apjl, 751, L21

\bibitem[{{Klimchuk}(2015)}]{Klimchuk:2015aa}
{Klimchuk}, J.~A. 2015, Philosophical Transactions of the Royal Society of
  London Series A, 373, 20140256

\bibitem[{{Lemen} {et~al.}(2012){Lemen}, {Title}, {Akin}, {Boerner}, {Chou},
  {Drake}, {Duncan}, {Edwards}, {Friedlaender}, {Heyman}, {Hurlburt}, {Katz},
  {Kushner}, {Levay}, {Lindgren}, {Mathur}, {McFeaters}, {Mitchell}, {Rehse},
  {Schrijver}, {Springer}, {Stern}, {Tarbell}, {Wuelser}, {Wolfson}, {Yanari},
  {Bookbinder}, {Cheimets}, {Caldwell}, {Deluca}, {Gates}, {Golub}, {Park},
  {Podgorski}, {Bush}, {Scherrer}, {Gummin}, {Smith}, {Auker}, {Jerram},
  {Pool}, {Soufli}, {Windt}, {Beardsley}, {Clapp}, {Lang}, \&
  {Waltham}}]{Lemen:2012aa}
{Lemen}, J.~R., {Title}, A.~M., {Akin}, D.~J., {et~al.} 2012, \solphys, 275, 17

\bibitem[{{Louis} {et~al.}(2015){Louis}, {Ravindra}, {Georgoulis}, \&
  {K{\"u}ker}}]{Louis:2015aa}
{Louis}, R.~E., {Ravindra}, B., {Georgoulis}, M.~K., \& {K{\"u}ker}, M. 2015,
  \solphys, 290, 1135

\bibitem[{{Mumford} \& {Erd{\'e}lyi}(2015)}]{Mumford:2015ab}
{Mumford}, S.~J., \& {Erd{\'e}lyi}, R. 2015, \mnras, 449, 1679

\bibitem[{{Mumford} {et~al.}(2015){Mumford}, {Fedun}, \&
  {Erd{\'e}lyi}}]{Mumford:2015aa}
{Mumford}, S.~J., {Fedun}, V., \& {Erd{\'e}lyi}, R. 2015, \apj, 799, 6

\bibitem[{{Nordlund} {et~al.}(2009){Nordlund}, {Stein}, \&
  {Asplund}}]{Nordlund:2009aa}
{Nordlund}, {\AA}., {Stein}, R.~F., \& {Asplund}, M. 2009, Living Reviews in
  Solar Physics, 6, doi:10.12942/lrsp-2009-2

\bibitem[{{November} \& {Simon}(1988)}]{November:1988aa}
{November}, L.~J., \& {Simon}, G.~W. 1988, \apj, 333, 427

\bibitem[{{Palacios} {et~al.}(2012){Palacios}, {Balmaceda}, {Dom{\'{\i}}nguez},
  {Cabello}, \& {Domingo}}]{Palacios:2012aa}
{Palacios}, J., {Balmaceda}, L.~A., {Dom{\'{\i}}nguez}, S.~V., {Cabello}, I.,
  \& {Domingo}, V. 2012, in Astronomical Society of the Pacific Conference
  Series, Vol. 454, Hinode-3: The 3rd Hinode Science Meeting, ed. T.~{Sekii},
  T.~{Watanabe}, \& T.~{Sakurai}, 51

\bibitem[{{Park} {et~al.}(2016){Park}, {Tsiropoula}, {Kontogiannis},
  {Tziotziou}, {Scullion}, \& {Doyle}}]{Park:2016aa}
{Park}, S.-H., {Tsiropoula}, G., {Kontogiannis}, I., {et~al.} 2016, \aap, 586,
  A25

\bibitem[{{Parker}(1972)}]{Parker:1972aa}
{Parker}, E.~N. 1972, \apj, 174, 499

\bibitem[{{Parker}(1983{\natexlab{a}})}]{Parker:1983ab}
---. 1983{\natexlab{a}}, \apj, 264, 642

\bibitem[{{Parker}(1983{\natexlab{b}})}]{Parker:1983aa}
---. 1983{\natexlab{b}}, \apj, 264, 635

\bibitem[{Pizer {et~al.}(1987)Pizer, Amburn, Austin, Cromartie, Geselowitz,
  Greer, ter Haar~Romeny, Zimmerman, \& Zuiderveld}]{pizer1987adaptive}
Pizer, S.~M., Amburn, E.~P., Austin, J.~D., {et~al.} 1987, Computer vision,
  graphics, and image processing, 39, 355

\bibitem[{{Scharmer} {et~al.}(2003){Scharmer}, {Bjelksjo}, {Korhonen},
  {Lindberg}, \& {Petterson}}]{Scharmer:2003aa}
{Scharmer}, G.~B., {Bjelksjo}, K., {Korhonen}, T.~K., {Lindberg}, B., \&
  {Petterson}, B. 2003, in \procspie, Vol. 4853, Innovative Telescopes and
  Instrumentation for Solar Astrophysics, ed. S.~L. {Keil} \& S.~V. {Avakyan},
  341--350

\bibitem[{{Scharmer} {et~al.}(2008){Scharmer}, {Narayan}, {Hillberg}, {de la
  Cruz Rodriguez}, {L{\"o}fdahl}, {Kiselman}, {S{\"u}tterlin}, {van Noort}, \&
  {Lagg}}]{Scharmer:2008aa}
{Scharmer}, G.~B., {Narayan}, G., {Hillberg}, T., {et~al.} 2008, \apjl, 689,
  L69

\bibitem[{{Shelyag} {et~al.}(2012){Shelyag}, {Fedun}, {Erd{\'e}lyi}, {Keenan},
  \& {Mathioudakis}}]{Shelyag:2012aa}
{Shelyag}, S., {Fedun}, V., {Erd{\'e}lyi}, R., {Keenan}, F.~P., \&
  {Mathioudakis}, M. 2012, in Astronomical Society of the Pacific Conference
  Series, Vol. 463, Second ATST-EAST Meeting: Magnetic Fields from the
  Photosphere to the Corona., ed. T.~R. {Rimmele}, A.~{Tritschler},
  F.~{W{\"o}ger}, M.~{Collados Vera}, H.~{Socas-Navarro}, R.~{Schlichenmaier},
  M.~{Carlsson}, T.~{Berger}, A.~{Cadavid}, P.~R. {Gilbert}, P.~R. {Goode}, \&
  M.~{Kn{\"o}lker}, 107

\bibitem[{{Shelyag} {et~al.}(2011){Shelyag}, {Keys}, {Mathioudakis}, \&
  {Keenan}}]{Shelyag:2011aa}
{Shelyag}, S., {Keys}, P., {Mathioudakis}, M., \& {Keenan}, F.~P. 2011, \aap,
  526, A5

\bibitem[{{Steiner} {et~al.}(2010){Steiner}, {Franz}, {Bello Gonz{\'a}lez},
  {Nutto}, {Rezaei}, {Mart{\'{\i}}nez Pillet}, {Bonet Navarro}, {del Toro
  Iniesta}, {Domingo}, {Solanki}, {Kn{\"o}lker}, {Schmidt}, {Barthol}, \&
  {Gandorfer}}]{Steiner:2010aa}
{Steiner}, O., {Franz}, M., {Bello Gonz{\'a}lez}, N., {et~al.} 2010, \apjl,
  723, L180

\bibitem[{{van Noort} {et~al.}(2005){van Noort}, {Rouppe van der Voort}, \&
  {L{\"o}fdahl}}]{van-Noort:2005aa}
{van Noort}, M., {Rouppe van der Voort}, L., \& {L{\"o}fdahl}, M.~G. 2005,
  \solphys, 228, 191

\bibitem[{{van Noort} \& {Rouppe van der Voort}(2008)}]{van-Noort:2008aa}
{van Noort}, M.~J., \& {Rouppe van der Voort}, L.~H.~M. 2008, \aap, 489, 429

\bibitem[{{Vargas Dom{\'{\i}}nguez} {et~al.}(2011){Vargas Dom{\'{\i}}nguez},
  {Palacios}, {Balmaceda}, {Cabello}, \& {Domingo}}]{Vargas-Dominguez:2011aa}
{Vargas Dom{\'{\i}}nguez}, S., {Palacios}, J., {Balmaceda}, L., {Cabello}, I.,
  \& {Domingo}, V. 2011, \mnras, 416, 148

\bibitem[{{Verma} {et~al.}(2013){Verma}, {Steffen}, \& {Denker}}]{Verma:2013aa}
{Verma}, M., {Steffen}, M., \& {Denker}, C. 2013, \aap, 555, A136

\bibitem[{{Wedemeyer} {et~al.}(2013){Wedemeyer}, {Scullion}, {Rouppe van der
  Voort}, {Bosnjak}, \& {Antolin}}]{Wedemeyer:2013aa}
{Wedemeyer}, S., {Scullion}, E., {Rouppe van der Voort}, L., {Bosnjak}, A., \&
  {Antolin}, P. 2013, \apj, 774, 123

\bibitem[{Wedemeyer-B\"ohm {et~al.}(2012)Wedemeyer-B\"ohm, Scullion, Steiner,
  van~der Voort, de~La~Cruz~Rodriguez, Fedun, \&
  Erd\'elyi}]{wedemeyer2012magnetic}
Wedemeyer-B\"ohm, S., Scullion, E., Steiner, O., {et~al.} 2012, Nature, 486,
  505

\bibitem[{{Withbroe} \& {Noyes}(1977)}]{Withbroe:1977aa}
{Withbroe}, G.~L., \& {Noyes}, R.~W. 1977, \araa, 15, 363

\end{thebibliography}

\end{document}